\documentclass{ws-ijmpe}
\begin{document}

\markboth{A. B. Balantekin, T. Dereli and Y. Pehlivan}
{Exactly Solvable Pairing Model}

\catchline{}{}{}{}{}

\title{EXACTLY SOLVABLE PAIRING MODEL USING AN EXTENSION OF
RICHARDSON-GAUDIN APPROACH}

\author{\footnotesize  A. B. BALANTEKIN}
\address{Department of Physics, University of Wisconsin\\
Madison, Wisconsin 53706 USA\\
baha@physics.wisc.edu }

\author{\footnotesize  T. DERELI}
\address{Department of Physics, Ko\c{c} University\\
34450 Sar{\i}yer, Istanbul, Turkey\\
tdereli@ku.edu.tr}

\author{\footnotesize  Y. PEHLIVAN}
\address{Department of Mathematics, Izmir University of Economics\\
35330 Bal\c{c}ova/IZMIR, Turkey\\
yamac@physics.wisc.edu}

\maketitle

\begin{history}
\received{\today}
%\revised{(revised date)}
%\accepted{(Day Month Year)}
%\comby{(xxxxxxxxxx)}
\end{history}

\begin{abstract}
We introduce a new class of exactly solvable boson pairing
models using the technique of Richardson and Gaudin.
Analytical expressions for all energy eigenvalues and first few
energy eigenstates are given. In addition, another solution to
Gaudin's equation is also mentioned. A relation with the
Calogero-Sutherland model is suggested.
\end{abstract}

\section{Introduction}

The notion of pairing plays a central role in the BCS
theory of superconductivity.\cite{BCS}
In 1963, R.W. Richardson showed that exact energy
eigenvalues and eigenstates of
the BCS pairing Hamiltonian can be computed if one can solve
a given set of highly coupled nonlinear
equations.\cite{Richardson1,Richardson2}
The limit of these equations in which the BCS coupling constant is very large
were also obtained by Gaudin in 1976.
Gaudin`s tool was the algebraic Bethe ansatz method.\cite{Gaudin1,Gaudin2}
Gaudin's Hamiltonians are the constants of motion of BCS Hamiltonian in the
large coupling constant limit.

In this paper, we present a new class of exactly solvable boson pairing
models using the technique pioneered by Richardson and Gaudin.
The technique is outlined in the next section and in Section 3 we
introduce the new class of exactly solvable models together with all
energy eigenvalues and first few energy eigenstates.

Richardson-Gaudin technique is based on an equation of which three solutions
are known. To each solution there is a corresponding set of mutually commuting
Hamiltonians and a related Lie algebra.
In Section 4 we present another solution to
Gaudin's equation. We identify the
corresponding set of mutually commuting Hamiltonians
together with the related Lie algebra.

\section{Gaudin Algebra}

Gaudin starts with the following operators\cite{Gaudin1}:
\begin{equation}\label{hGaudin}
h^{(0)}_i=\sum_{{{j=1}\atop{j\neq i}}}^N \sum_{\alpha=1}^3
w_{ij}^\alpha t_i^\alpha t_j^\alpha.
\end{equation}
Here $w_{ij}^\alpha$ are arbitrary complex numbers and
$t_i^\alpha$ satisfy the $SU(1,1)$ commutators:
\begin{equation}
\label{su2} [t_i^+,t_j^-]=-2\delta_{ij} t_j^0,\quad [t_i^0,t_j^\pm
]=\pm \delta_{ij}t_j^\pm.
\end{equation}
Then he shows that if we require these Hamiltonians to commute
\begin{equation}\label{hcommute}
[h^{(0)}_i,h^{(0)}_j]=0
\end{equation}
the coefficients $w_{ij}^\alpha$ must obey the following equation:
\begin{equation}\label{Gaudin's Equation}
w_{ij}^\alpha w_{jk}^\gamma + w_{ji}^\beta w_{ik}^\gamma -
w_{ik}^\alpha w_{jk}^\beta = 0.
\end{equation}
Assuming \textbf{i)} the coefficients are antisymmetric under the exchange
of $i$ and $j$
\begin{equation}\label{antisymmetry}
w_{ij}^\alpha+w_{ji}^\alpha=0
\end{equation}
\textbf{ii)} the coefficient $w_{ij}^\alpha$ can be expressed as a
function of the difference between two real parameters $u_i$
and $u_j$, and
\textbf{iii)} the operator $\sum_i t_i^0$ commutes with
$h^{(0)}_j$ for all $j$,
Gaudin found three solutions to Eq. (\ref{Gaudin's
Equation}). These solutions are given by
\begin{equation}\label{rationalsolution}
w_{ij}^\alpha = \frac{1}{u_i-u_j} \quad\mbox{for}\quad
\alpha=1,2,3 ,
\end{equation}
\begin{equation}\label{trigonometricsolution}
w_{ij}^0=p\cot [p(u_i-u_j)] \quad
w_{ij}^{1,2}=\frac{p}{\sin[p(u_i-u_j)]} ,
\end{equation}
\begin{equation}\label{hyperbolicsolution}
w_{ij}^0=p\coth[p(u_i-u_j)] \quad
w_{ij}^{1,2}=\frac{p}{\sinh[p(u_i-u_j)]} .
\end{equation}
The operators which are obtained by substituting these solutions
in Eq. (\ref{hGaudin}) are referred to as rational,
trigonometric and hyperbolic Gaudin magnet Hamiltonians,
respectively.\cite{Gaudin2,ush} In the rest of this section we
consider the rational Hamiltonian
\begin{equation}\label{DEFINE_h_i}
h^{(0)}_i=\sum_{{{j=1}\atop{j\neq i}}}^N
\frac{\overrightarrow{t}_i \cdot \overrightarrow{t}_j}{u_i-u_j} .
\end{equation}
The associated algebra, rational Gaudin algebra is an infinite
dimensional complex Lie algebra
generated by a one-parameter set of operators
$S^+(\lambda), S^-(\lambda)$ and $S^0(\lambda)$.
Commutators for rational Gaudin algebra are given as follows:
\begin{equation}\label{C1}
[S^+(\lambda),S^-(\mu)]=-2
\frac{S^0(\lambda)-S^0(\mu)}{\lambda-\mu}
\end{equation}
\begin{equation}
[S^0(\lambda),S^{\pm}(\mu)]=
\pm\frac{S^{\pm}(\lambda)-S^{\pm}(\mu)}{\lambda-\mu}
\end{equation}
\begin{equation}\label{C3}
[S^0(\lambda),S^0(\mu)]=[S^{\pm}(\lambda),S^{\pm}(\mu)]=0
\end{equation}
Commutators of these generators for $\lambda=\mu$
are found by taking the limit of
the right hand sides of the above commutators as $\lambda\to\mu$.

Let us define the operator
\begin{equation}\label{defineH}
H(\lambda)=S^0(\lambda)S^0(\lambda)-\frac{1}{2}S^+(\lambda)S^-(\lambda)-
                                    \frac{1}{2}S^-(\lambda)S^+(\lambda) .
\end{equation}
Using the commutators (\ref{C1}-\ref{C3}),
it is easy to show that $H(\lambda)$ forms a one parameter family
of mutually commuting operators:
\begin{equation}\label{HCommutators}
[H(\lambda),H(\mu)]=0.
\end{equation}
One can diagonalize these operators simultaneously, starting from a lowest
weight vector and using $S^+(\lambda)$ as step operators. Lowest
weight vector $|0>$ by satisfies
\begin{equation}\label{LOWEST_WEIGTH_STATE}
S^-(\lambda)|0>=0,\quad\mbox{and}\quad
S^0(\lambda)|0>=W(\lambda)|0>
\end{equation}
for every $\lambda\in\mathbb{C}$. Here $W(\lambda)$ is a
complex function. One gets
\begin{equation}
H(\lambda)|0>=E_0(\lambda)|0>, \>\>
E_0(\lambda)=W(\lambda)^2-W'(\lambda).
\end{equation}
We write the Bethe ansatz state
\begin{equation}\label{Bethe ansatz state}
|\xi_1,\xi_2,\dots,\xi_n>\equiv S^+(\xi_1) S^+(\xi_2)\dots
S^+(\xi_n)|0>
\end{equation}
for $n$ arbitrary complex numbers $\xi_1,\xi_2,\dots, \xi_n\in{\mathbb{C}}$.
This is an eigenvector of $H(\lambda)$ if these complex numbers
satisfy the following set of $n$ Bethe ansatz equations\cite{Gaudin2,ush}:
\begin{equation}\label{Bethe ansatz equations}
W(\xi_\alpha) = {\sum_{\beta=1, \beta \neq \alpha }^n}
\frac{1}{\xi_\alpha-\xi_\beta} \quad \mbox{for} \quad
\alpha=1,2,\dots,n.
\end{equation}
If $\xi_1,\xi_2,\dots, \xi_n\in{\mathbb{C}}$ is a solution of
above equations then (\ref{Bethe ansatz state}) is an eigenvector
of $H(\lambda)$ with the eigenvalue
\begin{equation}\label{Energy}
E_n(\lambda)=E_0(\lambda)-2\sum_{\alpha=1}^n
\frac{W(\lambda)-W(\xi_\alpha)}{\lambda-\xi_\alpha} .
\end{equation}

There exists a realization of the rational Gaudin algebra in terms
of the SU(1,1) generators of Eq. (\ref{su2}), given by
\begin{equation} \label{REALIZATION}
S^{0}(\lambda) =
\sum_{i=1}^N\frac{t_i^0}{u_i-\lambda}\quad\mbox{and}\quad
S^{\pm}(\lambda)=\sum_{i=1}^N \frac{t_i^\pm}{u_i-\lambda}.
\end{equation}
Here $u_1,u_2,\dots,u_N$ are arbitrary real numbers which are all
different from each other and $N \geq 0$. In
this realization $H(\lambda)$ of Eq. (\ref{defineH}) is given by
\begin{equation}
H(\lambda)=
\sum_{i,j=1}^N\frac{\overrightarrow{t_i}\cdot\overrightarrow{t_j}}
                        {(u_i-\lambda)(u_j-\lambda)}.
\end{equation}
$H(\lambda)$ has simple poles on the real axis.
Residues of $-H(\lambda)/2$ at the points $\lambda=u_i$ are
the rational Gaudin magnet Hamiltonians given in Eq.
(\ref{DEFINE_h_i}).
Eq. (\ref{HCommutators}) implies
\begin{equation}
[H(\lambda),h^{(0)}_i]=0, \quad\quad\mbox{and}\quad\quad
[h^{(0)}_i,h^{(0)}_j]=0.
\end{equation}

\section{A Model of Interacting Bosons}

The rational Gaudin algebra (\ref{C1}-\ref{C3}) can be realized in
terms of the boson operators:
\begin{equation}\label{R1}
S^{+}(\lambda)=\frac{1}{2}\sum_{\beta=1}^{n_1}\Lambda_\beta
\frac{b_\beta^\dagger b_\beta^\dagger}{x_1-\lambda}%
+\frac{1}{2}\sum_{\alpha=1}^{n_2}\Xi_\alpha%
\frac{a_\alpha^\dagger a_\alpha^\dagger}{x_2-\lambda}
\end{equation}
\begin{equation}
S^{-}(\lambda)=\frac{1}{2}\sum_{\beta=1}^{n_1}\Lambda_\beta %
\frac{b_\beta b_\beta}{x_1-\lambda}%
+\frac{1}{2}\sum_{\alpha=1}^{n_2}\Xi_\alpha%
\frac{a_\alpha a_\alpha}{x_2-\lambda}
\end{equation}
\begin{equation}\label{R3}
S^{0}(\lambda)=\frac{1}{4}\sum_{\beta=1}^{n_1}\frac{
b_\beta^\dagger b_\beta + b_\beta b_\beta^\dagger}{x_1-\lambda}%
+\frac{1}{4}\sum_{\alpha=1}^{n_2}\frac{a_\alpha^\dagger a_\alpha +
a_\alpha a_\alpha^\dagger}{x_2-\lambda}
\end{equation}
Here $x_1$ and $x_2$ are two real parameters which differ from
each other and $\Lambda_\mu^2=\Xi_\alpha^2=1$. There are two types
of bosons named $b_\beta$ and $a_\alpha$ for
$\beta=1,2,\dots,n_1$ and $\alpha=1,2,\dots,n_2$. The operators
$b_\beta,b_\beta^\dagger$ and $a_\alpha,a_\alpha^\dagger$ create or
annihilate these bosons:
\begin{equation}
[b_\beta, b_\gamma^\dagger]=\delta_{\beta \gamma}\quad
[a_\alpha,a_\sigma^\dagger]=\delta_{\alpha \sigma}.
\end{equation}
If we choose $n_1=5$, $n_2=1$, $\Lambda_\beta = 1$, and, $\Xi_\alpha =
-1$ then above operators
can be considered as a generalization of the $sd$-version of the
interacting boson model.
\cite{Armia:1976ky,Arima2,Balantekin:2003zf,Balantekin:2004yf}

The
lowest weight state is the boson vacuum and
corresponding $W(\lambda)$ is
\begin{equation}\label{W_FOR_BOSONS}
W(\lambda) =
\frac{1}{4}\left(\frac{n_1}{x_1-\lambda}+\frac{n_2}{x_2-\lambda}\right).
\end{equation}
In this realization, $H(\lambda)$ defined in Eq. (\ref{defineH}) is a
boson pairing Hamiltonian.

For simplicity we subtract the ground state energy and write down
$\tilde{H}(\lambda)=H(\lambda)-E_0(\lambda)$:
\begin{eqnarray}\label{PAIRING_HAMILTONIAN}
\tilde{H}(\lambda)&=\frac{1}{4} \left[ \frac{\hat{N}_d}{x_1
-\lambda} + \frac{\hat{N}_s}{x_2-\lambda} \right]^2 - \frac{1}{2}
\left[ \frac{\hat{N}_d}{(x_1 -\lambda)^2} -
\frac{\hat{N}_s}{(x_2-\lambda)^2} \right]\nonumber\\
&+W(\lambda) \left[ \frac{{\hat N}_d}{x_1 -\lambda} + \frac{{\hat
N}_s}{x_2-\lambda} \right] - S^+(\lambda) S^-(\lambda).
\end{eqnarray}
Here $\hat{N}_d$ and $\hat{N}_s$ are total number operators for
the $d$ and $s$ type bosons, i.e.
\begin{equation}
\hat{N}_d=\sum_{\mu} d_\mu^\dagger d_\mu
\quad\mbox{and}\quad%
\hat{N}_s= s^\dagger s .
\end{equation}
The last term in (\ref{PAIRING_HAMILTONIAN}) is a pairing operator
which corresponds to the $P_6$ pairing operator of the $sd$-boson
model. It involves the terms $d_\mu^\dagger d_\mu^\dagger d_\nu
d_\nu$ and $s^\dagger s^\dagger s s$ as
well as the cross terms $d_\mu^\dagger d_\mu^\dagger s
s$ and $s^\dagger s^\dagger d_\mu d_\mu$.
Therefore $\tilde{H}(\lambda)$ allows $s$-pairs to annihilate each
other to form $s$-pairs or $d$-pairs. Similarly $d$-pairs can also
annihilate each other to produce $s$ or $d$ pairs.

Since $\tilde{H}(\lambda)=H(\lambda)-E_0(\lambda)$,
energy eigenvalues of $\tilde{H}(\lambda)$ can obtained from those of
$H(\lambda)$  by subtracting the ground
state energy. Eigenvalues of $H(\lambda)$ were presented
in the previous section. Gaudin's equation (\ref{Energy}) tells us that
the energy eigenvalues of $\tilde{H}(\lambda)$ are
\begin{equation}\label{Energytilde}
\tilde{E}_n(\lambda)=-2\sum_{\alpha=1}^n
\frac{W(\lambda)-W(\xi_\alpha)}{\lambda-\xi_\alpha}.
\end{equation}
Here, $\xi_1,\xi_2,\dots, \xi_n\in{\mathbb{C}}$ are obtained by solving
the Bethe ansatz Eqs. (\ref{Bethe ansatz equations}).

Here we present a method to calculate these energy eigenvalues
in an analytical way without solving the Bethe ansatz equations. Let us start
with substituting $W(\lambda)$ given in
(\ref{W_FOR_BOSONS}) into the energy formula (\ref{Energytilde}) to find
\begin{equation}\label{Energytilde2}
\tilde{E}_n(\lambda)=-
2\sum_{\alpha=1}^n \left[\frac{-s_1}{(x_1-\lambda)(x_1 -
\xi_\alpha)}+\frac{-s_2}{(x_2-\lambda)(x_2-\xi_\alpha)}\right].
\end{equation}
We see that it is not necessary to compute all the unknowns
$\xi_1,\xi_2,\dots,\xi_n\in{\mathbb{C}}$ one by one to evaluate
$\tilde{E}_n(\lambda)$. It is sufficient to compute the
following sums:
\begin{equation}
S_1=\sum_{\alpha=1}^n \frac{1}{x_1 - \xi_\alpha}
\quad\quad\quad\quad
S_2=\sum_{\alpha=1}^n \frac{1}{x_2 - \xi_\alpha}.
\end{equation}
We compute $S_1$ and $S_2$ using the symmetries of the
equations of Bethe ansatz.
It is easy to show that Eqs. (\ref{Bethe ansatz equations})  imply
\begin{equation}\label{Sums}
\sum_{\alpha=1}^n W(\xi_\alpha) = 0 \quad\quad\mbox{and}\quad\quad
\sum_{\alpha=1}^n \xi_\alpha W(\xi_\alpha)=\frac{n(n-1)}{2}
\end{equation}
no matter what the form of $W(\lambda)$ is. Substituting
$W(\lambda)$ given in (\ref{W_FOR_BOSONS}) in above equalities one
finds two equations in $S_1$ and $S_2$. Solving these equations for
$S_1$ and $S_2$ and then substituting them in equation (\ref{Energytilde2})
one finds\cite{Balantekin:2004yf}
\begin{equation}\label{ENERGY}
\tilde{E}_n(\lambda)=
\frac{1}{(x_1-\lambda)(x_2-\lambda)}\left[n(n-1)+(n_1+n_2)\frac{n}{2}\right].
\end{equation}
This is an analytical expression for all energy eigenvalues of
$\tilde{H}(\lambda)$. It is computed without solving the
Bethe ansatz equations. To find the corresponding energy eigenstates, however,
one needs to solve the Bethe ansatz equations because each one of the
quantities $\xi_1,\xi_2,\dots,\xi_n\in{\mathbb{C}}$ is needed
in order to write down the Bethe ansatz state in equation (\ref{Bethe
ansatz state}).

Here, we present a method to turn the problem of solving the Bethe
ansatz equations
for the $n^{\mbox{th}}$ excited state
into a problem of finding the roots of an $n^{\mbox{th}}$ order polynomial.
We substitute
(\ref{W_FOR_BOSONS}) in the equations of Bethe ansatz and then
apply the change of variables $\xi_{\alpha} = x_2 + \zeta_{\alpha}
(x_1 - x_2)$. Bethe ansatz equations then assume the following form:
\begin{equation}\label{Bethe ansatz equations 2}
\sum_{ \beta\neq\alpha }^n \frac{1}{\zeta_\alpha-\zeta_\beta} +
\frac{n_2/4}{\zeta_{\alpha}} -\frac{n_1/4}{1-\zeta_{\alpha}} = 0
\quad \mbox{for} \quad \alpha=1,2,\dots,n.
\end{equation}
It was shown by Stieltjes\cite{stiel} that when $\zeta_\alpha$
obey (\ref{Bethe ansatz equations 2}), the polynomial
\begin{equation}
p_n(\zeta) = \prod_{\alpha =1}^{n} (\zeta - \zeta_{\alpha})
\end{equation}
satisfies the hypergeometric differential equation
\begin{equation}
\zeta(1-\zeta) p_n''(\zeta) -
\frac{n_1}{2} p_n'(\zeta) + n(n+\frac{n_1+n_2}{2}-1)
p_n (\zeta) = 0.
\end{equation}
This means that $\zeta_{\alpha}$ for $\alpha=1,2,\dots,n$ are the
roots of the following polynomial:
\begin{equation}\label{Polynomial}
\sum_{k=1}^\infty \frac{(-n)_k (n+n_1/2+n_2/2-1)_k}{k!(n_2/2)_k }
\zeta^k=0.
\end{equation}
\begin{eqnarray}
{\rm Here} \>\>\>(a)_0&=&0 \\
(a)_n &=& a (a+1) \cdots (a+n-1)\quad \mbox{for}\quad
n=1,2,\cdots
\end{eqnarray}
The polynomial in (\ref{Polynomial}) terminates
when $k=n$ because $(-n)_{n+1}=0$. Consequently, finding the solution
of the equations of Bethe ansatz for the $n^{th}$ energy level is now
reduced to the problem of finding the roots on an $n^{th}$ order
polynomial. In principle, we can find the roots of polynomials analytically
up to the forth order. Therefore, first four excited energy eigenstates
can be computed analytically. Solutions for the higher order eigenstates
can be performed using numerical techniques.

\section{A New Solution to Gaudin's Equation}

In this section we briefly mention another solution to Gaudin's equation
given in (\ref{Gaudin's Equation}). Instead of the
constraint
(i) given in Eq. (\ref{antisymmetry})
this new solution obeys
$w_{ij}^\alpha+w_{ji}^\alpha=2q$
together with the other two constraints (ii) and (iii).
The solution is given by\cite{Balantekin3}
\begin{equation}\label{newsolution}
w_{ij}^\alpha = q\coth[q(u_i-u_j)]+q \quad \mbox{for}\quad
\alpha=1,2,3.
\end{equation}
Here $q$ is a complex parameter. In the limit where $q\to 0$ this
new solution approaches to the rational solution of Gaudin given
by (\ref{rationalsolution}). For this reason we will denote the
corresponding Gaudin magnet operators by $h^{(q)}_i$. In other words
\begin{equation}\label{hnew}
h^{(q)}_i=\sum_{{{j=1}\atop{j\neq i}}}^N
[q\coth[q(u_i-u_j)]+q]\overrightarrow{t}_i \cdot
\overrightarrow{t}_j.
\end{equation}
These operators mutually commute: $[h^{(q)}_i,h^{(q)}_j]=0$. This
Hamiltonian and its Bethe ansatz solution was discussed in Ref. 12.

\section{A Connection with Calogero-Sutherland Model}

Calogero model was first introduced in 1969 as a many-body system
in one dimension with inverse square two-body
interactions.\cite{Calogero}
A variant of Calogero model introduced by B. Sutherland in 1971
includes two-body inverse sine square potentials.\cite{Sutherland}
Both models are prime examples of exactly solvable systems and have many
interesting features. Originally both models involved scalar particles.
But generalizations of these models to particles with
spin degree of freedom which are called spin-Calogero
and spin-Sutherland models also received increasing
attention. \cite{Gibbons,Wojciechowski,Ha,Kawakami,Polychronakos}
Spin-Calogero model Hamiltonian is given by
\begin{equation}\label{SPIN_CALOGERO_MODEL}
H^{(0)}=\sum_{i=1}^{N}\frac{1}{2}p_i+g\sum_{{i,j=1}\atop{i<j}}^{N}
\frac{\overrightarrow{t}_i\cdot\overrightarrow{t}_j}{\left(
x_{i}-x_{j}\right)^{2}}.
\end{equation}
This model describes non-relativistic particles which are free to
move in one-dimension. Here $x_{i}$ denote the positions, $p_{i}$
denote the momenta and masses are scaled to unity.
$\overrightarrow{t}_i$ for $i=1,2,\dots,N$ represent independent
angular momenta of the particles.

Spin-Sutherland model, on the other hand, is given with the Hamiltonian
\begin{equation}\label{Spin Sutherland Model}
H^{(q)}=\frac{1}{2}\sum_{i=1}^{N}p_i^{2}+
q^2g\sum_{{i,j=1}\atop{i<j}}^{N}
\frac{\overrightarrow{t}_i \cdot
\overrightarrow{t}_j}{\sinh^{2}[q(x_{i}-x_{j})]}.
\end{equation}
Here $q$ is a complex parameter such that in the limit where $q\to
0$, the later model approaches to the former. It is possible to
view Sutherland's model as a Calogero model on a circle, i.e. a
system of particles which are constrained to move on the circle
and interacting through Calogero potential along the cord distance
between them. If we adopt this view then $q=i\pi/d$ where $d$ is
the circumference of the circle.

One can write the spin-Calogero model Hamiltonian
using the rational Gaudin magnet Hamiltonians
given in Eq. (\ref{DEFINE_h_i}) as
\begin{equation}\label{Hamiltonian}
H^{(0)}=\frac{1}{2}\sum_{i=1}^N
\left(p_i^2+i\frac{g}{2}[p_i,h_i]\right).
\end{equation}
Similarly, the spin-Sutherland model Hamiltonian
can be written in the same form
\begin{equation}\label{qHamiltonian}
H^{(q)}=\frac{1}{2}\sum_{i=1}^N
\left(p_i^2+i\frac{g}{2}[p_i,h_i^{(q)}]\right)
\end{equation}
but this time we using the operators $h_i^{(q)}$ which are given by Eq.
(\ref{hnew}).

This connection between Calogero-Sutherland model and the Gaudin
algebras is closely related to the Lax formulation of
Calogero-Sutherland model. To see this let us first substitute
$w_{ij}^1=w_{ij}^2=w_{ij}^3=x(u_i-u_j)$ in Eq. \ref{Gaudin's
Equation}. Here $x(\xi)$ is a function which obeys
$x(\xi)+x(-\xi)=2q$. If we then take the derivative of Eq.
\ref{Gaudin's Equation} with respect to $u_i$ and substitute
$\xi=u_j-u_i$ and $\eta=u_i-u_k$ we find
\begin{equation*}
x(\xi)x^\prime(\eta)-x^\prime(\xi)x(\eta)=x(\xi+\eta)[x^\prime(\eta)-x^\prime(\xi)].
\end{equation*}
It is well known that when $x(\xi)$ satisfies this equation, a
many body system with the two-body potential $v=-x^\prime$ is a
Lax type system. In other words, its equations of motion can be
written in the form of a Lax pair (see Ref. 20 for a review). If
we use the rational solution of Eq. \ref{Gaudin's Equation} given
in Eq. \ref{rationalsolution} then the two-body potential
$v=x^\prime$ is equal to the Calogero potential. On the other
hand, if one uses of the new solution given by Eq.
\ref{newsolution} then $v=-x^\prime$ turn out to be the Sutherland
potential.

\section{Conclusion}

In this paper we suggested a few possible directions to extend the method of
Richardson and Gaudin. Results of Section 3 are very encouraging
because we see that one can obtain
analytical results without explicitly solving the equations of Bethe ansatz.
The method of transforming the equations of Bethe ansatz into a problem
of finding the roots of polynomials is also introduced. Finding
the roots of polynomials is much more established problem and this
trick may come handy in numerical calculations as well.

In this paper we relaxed one of the conditions imposed by Gaudin
on the solutions of Eq. (\ref{Gaudin's Equation}). This way we
obtained a new and physically interesting solution. A new set of
mutually commuting Hamiltonians $h_i^{(q)}$ are thus identified.
Evidently, these Hamiltonians can be related to spin-Sutherland
model in a way which is parallel to the relation between the
rational Gaudin magnet Hamiltonians and the spin-Calogero model.
Application of algebraic Bethe ansatz technique to $h_i^{(q)}$
yields a new Gaudin type algebra. This algebra admits a one
parameter family of Hamiltonians $H(\lambda)$ in a way which is
analogous to other Gaudin algebras.

\section*{Acknowledgments}

This  work   was supported in  part  by   the  U.S.  National
Science Foundation Grants No.\ INT-0352192 and PHY-0244384 at the
University of  Wisconsin, and  in  part by  the  University of
Wisconsin Research Committee   with  funds  granted by the
Wisconsin Alumni  Research Foundation. The research of YP was
supported in part by the Turkish Scientific and Technical
Research Council
(T{\"U}B{\.I}TAK).

\end{document}